\documentclass[%
 aip,
rsi,%
 amsmath,amssymb,
 reprint,%
]{revtex4-1}

\usepackage{graphicx}
\usepackage{dcolumn}
\usepackage{bm}
\usepackage{url}
\usepackage{notes2bib}

\begin{document}

\preprint{AIP/123-QED}

\title{FPGA based demodulation of laser induced fluorescence in plasmas}

\author{Sean W. Mattingly}
\author{Fred Skiff}%
\affiliation{ 
Department of Physics and Astronomy, University of Iowa, Iowa City, IA 52242 USA
}%

\date{\today}


\begin{abstract}

    We present a field programmable gate array (FPGA) based system that counts
    photons from laser induced fluorescence (LIF) on a laboratory plasma.  This
    is accomplished with FPGA based up / down counters that demodulate the
    data, giving a background-subtracted LIF signal stream that is updated with a new point
    as each laser amplitude modulation cycle completes.  We demonstrate using
    the FPGA to modulate a laser at 1 MHz and demodulate the resulting LIF data
    stream. This data stream is used to calculate an LIF - based measurement
    sampled at 1 MHz of a plasma ion fluctuation spectrum.

\end{abstract}

\pacs{}
\keywords{laser induced fluorescence, plasma, FPGA, fluctuation, correlation}
\maketitle

\section{\label{sec:level1} Introduction}

Laser induced fluorescence (LIF) is a spectroscopic diagnostic on various types
of plasmas including small laboratory plasmas upon which it was pioneered
\cite{Stern1975}, fusion tokamak divertors \cite{Gorbunov2010}, etching plasmas
\cite{Hershkowitz1997}, and more exotic plasmas \cite{Strickler, Anderegg1997}.
While LIF is primarily used as a diagnostic to find low order moments ($n$,
$\langle v \rangle$, $T_i$) of the plasma ion velocity distribution function,
it is also used for examining fundamental plasma physics through its ability to
nonperturbatively probe the ion distribution function. Fundamental plasma
measurements using LIF include local field measurements through optical tagging
techniques \cite{Stern1985}; motion of the plasma dielectric \cite{Skiff1987};
and, more recently, kinetic modes and fluctuations derived from correlation
functions \cite{Diallo2005}. These more complicated plasma measurements require
not only measurements of the plasma distribution function, but also background
subtracted measurements as a function of time. This is especially the case
with plasma fluctuation measurements.

Two major drawbacks plague LIF. First, the signal is limited due to Poisson
counting statistics.  Second, background collision induced fluorescence
threatens to flood the LIF signal. A variety of noise reduction techniques
circumvent these issues. We focus on a technique to remove background collision
induced fluorescence that has been used since the first demonstration of LIF:
amplitude modulation of the stimulating laser and demodulation of the resultant
modulated LIF signal.  Subtracting the demodulated components removes the
background collision induced fluorescence. 

Traditionally, this background subtraction is achieved with a lock-in amplifier
tuned to the modulation frequency of the laser. However, if one wants to
measure a laser induced fluorescence signal as a function of time, rather than
its strength as a function of wavelength, a lock-in is not useful and instead
digitization is warranted. Typically, digitization is accomplished by constant
fraction discriminators that convert the photon multiplier tube (PMT) pulses
into digital logic pulses. These digitized pulses are then histogrammed and
saved for later processing. Or, if discriminators are not available, then raw
voltages from the PMTs themselves are sampled and saved to be processed later.
Both the digital and analog methods require great cost in storage, acquisition
time, and post processing time. Moreover, the problem of removing the
background collision induced fluorescence still remains in both cases. 

However, there is a demodulation process that may be done at the time of
measurement for digitized photon pulses.  This technique\cite{Pelissier1996}
uses up / down counters in a digital logic circuit in order to remove the
background fluorescence at the time of measurement. It
accomplishes this by counting up while the laser is on and counting down while
the laser is off. At the end of each laser modulation cycle, it outputs the
result. Thus for each laser modulation cycle it obtains the background -
subtracted demodulated LIF photon count.

Since this technique is done entirely on digitized PMT signals, field
programmable gate arrays (FPGAs) are well suited to implement these up / down
counters for five reasons. First, counters do not take up many logic elements
on a typical FPGA, leaving ample room for more channels or in place digital
signal processing.  Second, the hardware development languages for FPGAs
support object instantiation so that numerous readout channels can be created
easily and in parallel. This is in contrast with the need, for example, to
solder and test 32 separate identical up / down circuits for a set of two 16
element PMTs.  Instead, this is all accomplished on a single chip. Third, FPGAs
are easily reconfigurable through just software and a cable - changes can be
made on the fly, without soldering or buying components.  Fourth, FPGAs are
clocked to an oscillator which is distributed throughout the circuit, easily
achieving synchronization between the laser and the up / down counters.
Finally, while examples abound of FPGAs seeing more use in larger plasma
experiments\cite{Naylor2010,Lovell2016,Balboa2010}, they are quickly becoming
cost effective enough and user friendly enough to be used in smaller
laboratories\cite{Schwettmann2011a,Ryou2017,Lusardi2017}.

An equivalent background subtraction method to that presented here using off
the shelf components uses a boxcar averager feeding into a lock - in
amplifier\cite{Rutledge1987}. This also results in a demodulated signal stream,
with a point added at the end of each laser modulation cycle.  Zero dead time
boxcars are now commercially available with performance matching or exceeding
what we present here. Faster speeds may also be obtained with fast digitizing
oscilloscopes, but these do not do the demodulation that we present here. In
either case, synchronizing this process over many PMTs is difficult as a large
number of physical devices are required, and any misalignment in
synchronization is twofold damaging to the result. 

What we seek, then, is to introduce a different strategy for acquiring
modulated LIF data across multiple arrays of PMTs. This strategy is made
possible by FPGAs. A single chip can acquire data that would take 32 sets of
boxcar / lock - in combinations. Moreover, the data acquired the way we present
is ready to be processed on the FPGA, a possibility that is missing from a
boxcar - lock in combination. This is particularly useful in applications that
require statistical averaging, since a running average will both reduce signal
to noise ratio and constrain the needed storage space to a constant.

In this Article, we present an FPGA based up / down counter LIF demodulation
scheme. We show how to synchronize the asynchronous digitized pulses to the
FPGA clock domain and the implementation of the up / down counters for multiple
parallel channels. This scheme demodulates the LIF data on a single device,
resulting in a digitized stream of LIF photon counts. This stream is ready for
time domain analysis, such as detecting plasma fluctuations through cross
correlations. We verify the FPGA based LIF demodulation on 32 data channels
from two 16 - element PMTs by comparing it to a traditional lock-in LIF
measurement. Finally, we show an example of an application of this method by
using the digitized LIF photon stream to calculate a cross correlation and
statistically average it with time.  This averaged cross correlation is used to
calculate a spectrum of a plasma.

This FPGA based demodulation scheme is important for any type of LIF
measurement that requires a digitized stream of background subtracted LIF
photons, since it removes a major component of measurement deadtime. It is
particularly useful in a measurement where statistical averaging is required to
achieve a good signal to noise ratio.

\section{System Design}

\begin{figure} \includegraphics{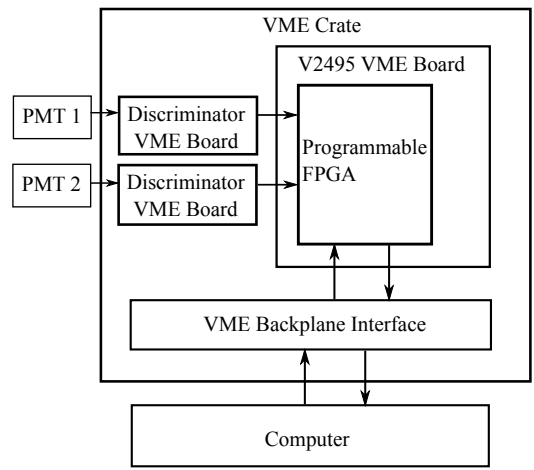} \caption{The high
        level system design. This shows the individual boards used in the VME
data acquisition system, and their connections to the PMTs and the computer. }
\label{fig:sys_design} \end{figure}

The system is housed in a Versa Module Europa bus (VMEbus) crate, a CAEN SpA
VME8002. VMEbus is a bus architecture that provides both housing for
electronics circuit boards and a communications interface connecting the boards
to each other through memory addressing. The pulse trains from two separate 16
element PMTs, Hamamatsu model H10515B-200, are fed into two 16 element falling
edge discriminators, CAEN SpA model V895, via coaxial cables with LEMO
connectors. The 16 element PMTs may be used for increased spatial resolution,
but here we simply sum them after the FPGA based demodulation to improve our
signal to noise ratio. The discriminators convert the PMT data pulses into ECL
logic pulses. The resulting ECL logic pulse train is sent via ribbon cable into
a VME board, CAEN Spa model V2495, housing a user programmable FPGA (UFPGA), an
Altera Cyclone V 5CGXBC4C6F27C7.  On the UFPGA, the data is demodulated, stored
in on - UFPGA memory, and sent to the computer through the VMEbus.  This high
level system design is shown in Fig.~\ref{fig:sys_design}.  A Struck GmbH
SIS3153 provides a USB interface to the computer to connect to the VMEbus.

\begin{figure}
    \includegraphics{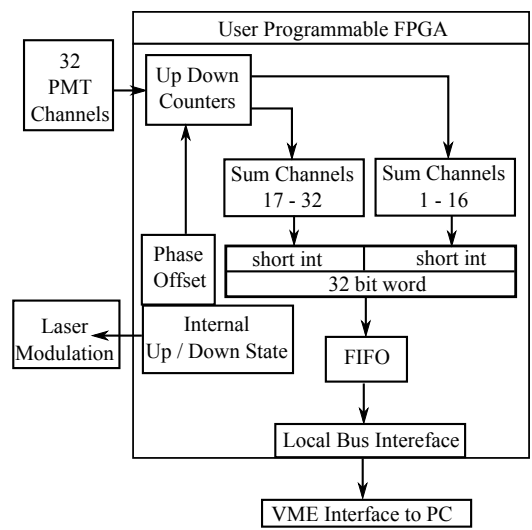}
    \caption{Logic on the user programmable FPGA. This logic handles three major tasks: input / output,
    the up / down counters, and storage of the demodulated LIF counts before it is read to the computer (FIFO).}
    \label{fig:ufpga_design}
\end{figure}

The logic on the UFPGA is laid out in Fig.~\ref{fig:ufpga_design}.  The 32
channel ECL logic pulses, 16 from each PMT / discriminator pair, are fed into
32 up / down counter channels on the UFPGA. A periodic internal up / down state,
created by a clock synchronized internal counter, governs whether the counters
count up or down. This signal is also fanned out with a user configurable phase
to a TTL output connector, providing a control signal for laser modulation.
This signal's - and thus the laser modulation's - frequency is either 1 MHz or
100 kHz. At each rising edge of this laser state the counters' values are
sampled and then reset to zero. Each sample from the 16 counters is summed and
the result drives an Altera FIFO input line \cite{FIFO}.  The FIFO output is
connected to a local bus interface, which interfaces with the VME backplane via
the V2495 hardware and is read out by the PC.

A finite state machine, not shown in Fig.~\ref{fig:ufpga_design}, controls the
status of the UFPGA, whether it is acquiring data, reading out the FIFO to the
PC via VME, or idling. The user changes the FSM state through adjusting control
register bits via VME memory write accesses.

\section{Up Down Counter Design}

The up / down counter requires a few intricacies. Firstly, the input pulses are
ECL logic pulses of width $5$ ns while the UFPGA clock is 50 MHz.
Secondly, these input pulses are not synchronized. Thus the problem is that one
needs to bring asynchronous 200 MHz clock signals into a 50 MHz clock domain.
This is solved using a Flancter\cite{Weinstein2000} circuit that has been
modified to count up or down based on an input status signal\cite{DOULOS}. The
Flancter uses a twisted pair of flip flops to set a status flag when an
incoming pulse is detected. The pulse needs to be long enough
to trip a flip flop - generally only 1 ns.

The flancter circuit is controlled by a finite state machine that requires four
clock cycles after a data pulse is latched to increment a counter. Pulses
that arrive during this state would be missed. This is another setback in the
process of counting these faster data pulses.

\begin{figure*}
    \includegraphics{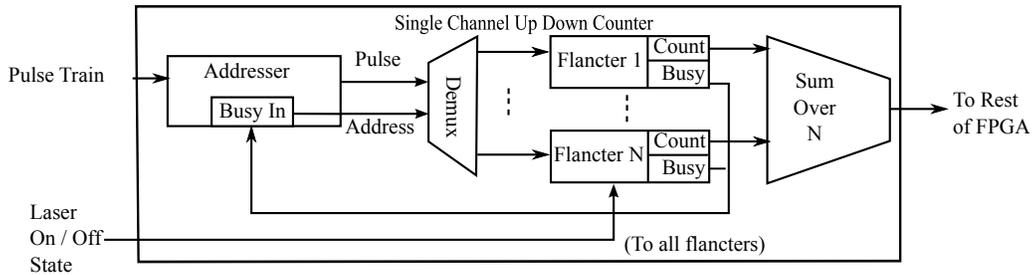}
    \caption{Block design of a single channel's up / down counter. Each incoming pulse is automatically routed
    to an idle flancter so that no pulse is missed.}
    \label{fig:up_down_design}
\end{figure*}

This problem is solved by taking advantage of the massively parallel nature of
FPGAs. Each up / down counter has a configurable (at compile time) number of
flancter circuits. Busy signals from occupied flancters are input to an
addresser in order to route the next pulse to an available flancter.
Since the user can scale the number of flancters as needed, the
bandwidth of the counting circuit is limited by FPGA signal propagation times.
This design is robust with respect to varying input data pulse lengths.
Fig.~\ref{fig:up_down_design} shows a diagram of the flancter busy
line addresser and demutiplexer on the input PMT pulse train.

\section{Experiment Setup}

\begin{figure} \includegraphics{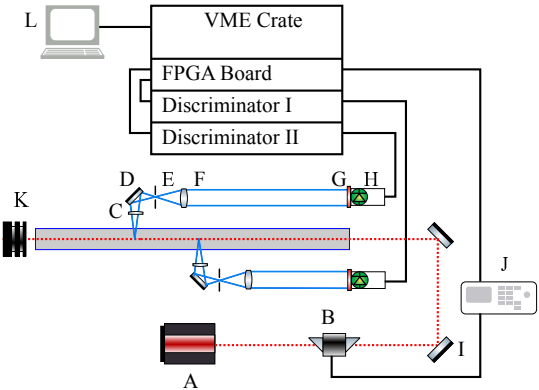} \caption{
                Experimental set up for acquiring LIF to test the FPGA. A -
                diode laser. B - electro optical modulator. C - aspheric lens.
                D - mirror. E - pinhole. F - collimating Fresnel lens. G - 442
                nm filter (1nm bandwidth) H - Hamamatsu H10515B-200 PMT.  I -
                mirrors. J - arbitrary function generator. K - beam dump. L -
                data acquisition PC. This setup can acquire streams
                of photons from two independent and moveable positions.
} \label{fig:setup} \end{figure}

The UFPGA is tested on a laboratory plasma device set up to obtain laser
induced fluorescence, shown in Fig.~\ref{fig:setup}. The plasma is a cylidrical
axially magnetized plasma column of singly ionized Argon (Ar II) created with
an RF inductively coupled source. Ion density and electron temperature are
measured with a Langmuir probe and are typically $n_i \approx 5e9$ cm$^{-3}$
and $T_e \approx 8$ eV. LIF traditionally applied shows the ion temperature
$T_i \approx 0.08$ eV. The plasma column is $230$ cm long with radius $\approx
2.5$ cm. Current carrying coils ensconce the chamber in a solenoid to create an
axial magnetic field of 667 Gauss. The plasma chamber also contains two
independently movable sets of light collecting optics focused on the axis of
the plasma.

These light collection optics gather light from the plasma which includes laser
induced fluorescence. They both spatially filter via a pinhole and spectrally
filter via optical filters centered near 442 nm. The laser power at the plasma
chamber point of entry is 30 mW - which is already close to
saturation broadening for our plasma.

The laser is a TOPTICA TA 100 Diode laser. Laser amplitude modulation is
achieved with a Conoptics electro-optic modulator model 390-2P. This laser
excites the Ar II metastable state 3s$^2$3p$^4$($^3$P)3d~$^4$F$_{7/2}$ with
$668$ nm light to the state 3s$^2$3p$^4$($^3$P)4p~$^4$D$_{5/2}^\circ$, which
then decays to 3s$^2$3p$^4$($^3$P)4p~$^4$P$_{3/2}^\circ$ while emitting light
near $442$ nm\cite{NIST2014}.  The laser wavelength is tunable across the
Zeeman split absorption spectrum of the $^4$F$_{7/2}$ $\rightarrow$
$^4$D$_{5/2}^\circ$ transition.  By tuning the laser wavelength, we can measure
the transition absorption spectrum that has been broadened by the ion velocity
distribution function of the plasma.

\section{Lock-In verification}

For verification, a lock-in is also connected to the discriminators. The V895
discriminators, in addition to outputting ECL logic pulses, also output a
single voltage signal proportional to the number of currently activated
discriminators. This signal is sent through an analog RC low pass filter with
RC $\approx$ 1 $\mu$s then into a Zurich Instruments MFLI5M lock-in amplifier.
The lock-in reference frequency is set to the laser modulation frequency.

\begin{figure} \includegraphics{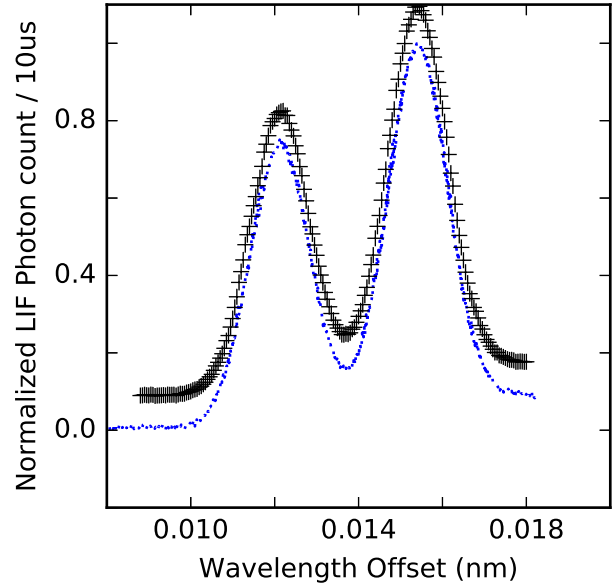} \caption{Lock-in data
                and FPGA absorption spectra. The dotted line denotes lock-in
                data and the plus ($+$) symbols correspond to FPGA data. The
                abscissa is the positive offset from $668.6$ nm. The two data sets
                have been normalized relative to each dataset's peak.  Note the
        zero of the FPGA data has been positively offset to make comparison
easier. Zeeman splitting of the transition is apparent in these curves.}
\label{fig:fpga_verify} \end{figure}

Two sets of data are taken to verify the system. The first we take by scanning
the laser across the entire absorption spectrum while reading the output from
the lock-in.  The lock-in internal RC time constant is slow relative to the
laser scanning speed. The second data set is taken with the UFPGA. At a given
wavelength, $6.4$ s of up / down data are taken and then averaged to give an
LIF rate. This data set length was chosen arbitrarily to ensure a high signal
to noise ratio.  This is repeated for wavelengths across the entire absorption
spectrum. The results are shown in Fig.~\ref{fig:fpga_verify}.

This measurement, while useful for verification since it recreates the output
of a lock-in amplifier, is time consuming and is not the strength of the UFPGA
setup. Indeed, when used this way, the UFPGA is slower and has worse resolution
than the lock-in.  The UFPGA setup's strength is an application that needs a
digitized stream (or streams), in time, of LIF photon counts. A lock-in
amplifier is not useful for obtaining this type of data.  We show a cross
correlation as an example of this kind of application.

\section{Example Application: Cross Correlation}

Cross correlation is a powerful tool for isolating and removing photon
statistics noise.  As an example, we provide a program which takes the
demodulated time series data from each of the two sets of counters, cross
correlates it, and takes a running pointwise average of the cross correlation
with earlier cross correlations.  By doing this, we have created a process that
finds the spectrum of plasma fluctuations by reducing the photon statistics
noise floor through time averaging. 

We choose cross correlation as an example because it demonstrates the strengths
of this FPGA based demodulation. There is no processing needed for
demodulation. The digitized LIF stream is directly used to calculate the cross
correlation. Successive cross correlations can be statistically averaged to
improve the signal to noise ratio. An application such as this that requires
statistical averaging is where the digitized data from this FPGA is
advantageous compared to a lock-in.

We accomplish this with a C program on the data acquisition computer. The
program first sets up the UFPGA with the desired parameters for the experiment.
It then lets the UFPGA acquire until it fills the onboard FIFO - 32k elements.
Then it downloads the time series demodulated LIF data through the VME
backplane.  With this LIF data set, it performs a cross correlation and adds
the result to a running average. Cross correlation is done in frequency space
with the assistance of the FFTW3 libraries\cite{Frigo1999}. With this
optimization it takes less time to perform the cross correlation than to
acquire data.  To take advantage of this, the acquisition loop is split into
two threads using the OpenMP library \cite{OPENMP}. Thread blocks are in place
to ensure that, when one thread is acquiring data, the other is calculating a
cross correlation of its already acquired data set. With this scheme, the only
data dead time is during data acquisition itself and transferring data to the
computer.

\begin{figure} \includegraphics{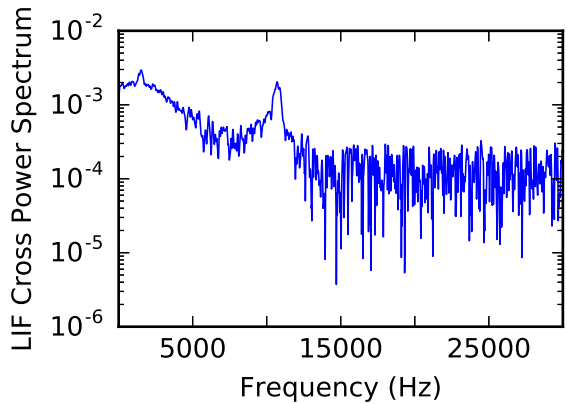} \caption{Gaussian windowed
                spectrum of the time averaged cross correlation of the
                demodulated LIF data streams. An ion acoustic peak at 1.2 kHz
                and a drift peak near 10 kHz are visible. Above this the plasma
        exhibits noise.} \label{fig:fpga_spec} \end{figure}

One minute of demodulated LIF data, in $32$ms groups with a $1$ MHz laser
modulation frequency, was acquired and averaged with this setup. $32$ ms groups
were taken since this is how long the 32K element FIFO on the UFPGA fills up at
1 MHz. An upper bound time to transfer this data from the UFPGA to the
computer is $10$ ms.  To see the fluctuation spectrum, the final statistically
averaged cross correlation function is point wise multiplied with a Gaussian
windowing function of width $5$ ms and Fourier transformed.  The resulting
spectrum is shown in Fig.~\ref{fig:fpga_spec}. A standing wave bounded mode in
the plasma chamber from the ion acoustic mode is visible at 1.2 kHz and a drift
wave peak is visible near $f^* \approx 10$kHz. 

A traditional VME histogramming scalar acquisition obtains a similar spectrum
from about 20 minutes of acquisition, writing about 5 gigabytes of raw data
that then must be demodulated, cross correlated, and statistically averaged
afterwards. Moreover, the laser modulation frequency is only 100 kHz. In
comparison, the FPGA and C program we present writes a 500 kB file containing
60 seconds worth of statistically averaged cross correlation after 1.5 minutes
with a 1 MHz laser modulation frequency.  While the dead time is not
eliminated, it has been drastically reduced.  With our particular set up, the
FPGA makes possible both the demodulation, and thus dead time reduction, and
the faster laser modulation frequency.

While this particular plasma only has noise at higher frequencies, we have
validated the FPGA's ability to find fluctuation spectra for higher sampling
frequency. This, combined with the removal of the dead time needed to demodulate
the LIF signal, means that this FPGA demodulation method is useful for finding
higher frequency plasma fluctuation spectra for low signal to noise ratio plasmas.

\begin{figure} \includegraphics{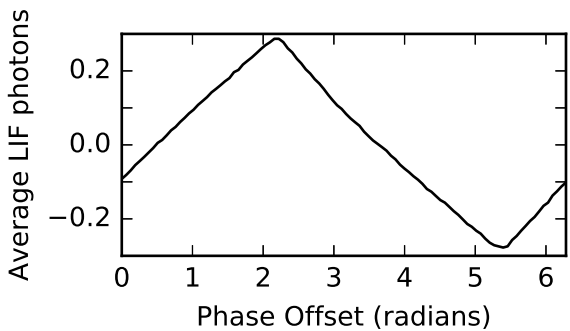} \caption{Average LIF signal
                per up / down counting cycle when the laser logic signal is
                phase shifted. The offset giving a maximum is chosen for a
                given setup. This can be adjusted when the setup changes. A 1
                MHz laser modulation frequency was used here. The triangle wave 
                verifies the up / down counters are counting LIF photons.}
                \label{fig:fpga_phase_scan} \end{figure}

\section{Summary and Future Work}

We presented a user programmable FPGA data acquisition system that gives a
stream of LIF photon counts that is updated as laser amplitude modulation cycle
completes. This system was verified against a lock-in measurement of the
absorption spectrum of a Zeeman split transition line in a singly charged Argon
ion magnetized plasma. We also presented a PC based C program for finding the
cross correlation of the demodulated LIF data.  Averaging this cross
correlation over time lowers the photon statistics noise floor, making more
detailed nonperturbative measurements of the ion density or temperature
fluctuation spectrum through LIF possible.

Future work should take greater advantage of the strengths of an FPGA.  This
includes moving the cross correlation, or another application that requires
time averaging, to the user programmable FPGA itself so that it removes the
UFPGA to PC data transfer bottleneck. Another possibility is integrating
impulse response filters in the FPGA logic, combining the strengths of a
lock-in amplifier with the strengths of a digitized stream of LIF photons. It
is possible that the dead time can be completely eliminated for these
applications, leaving only the time required to achieve a desirable signal to
noise ratio.

Finally, the FPGA makes spectral measurements possible at much higher
frequencies. A science avenue of future work involves observing high frequency
fluctuations with LIF closer to this new Nyquist frequency of 500 kHz. It is
also possible, by doubling the FPGA based phased-lock-loop output clock frequency,
to increase the laser modulation frequency to 2 MHz.

\section{Supplementary Material}

The VHDL code for the up / down counters, the C code for the cross correlation,
and supporting files for the V2495 board are provided as supplementary material
for this paper. Both a Quartus archive (.qar) of the entire project and a zip
file of only the VHDL files is included. The code for finding the correct laser
logic signal offset by scanning the relative phase through all values is also
included.  A typical phase scan is shown in Fig.~\ref{fig:fpga_phase_scan}.
This serves as another verification of the expected behavior of the up down
counters.  The entire FPGA project was written in and compiled using Quartus II 15.0
64 bit Web Edition, which is free, while the C code was written in Eclipse.

\section{Acknowledgements}

The first author wishes to acknowledge a donation of useful lab equipment from
Ingsheng Kung and the Kung family. This work is supported by the US DOE under
the NSF-DOE program with grant number DE-FG02-99ER54543.

%


\end{document}